\documentclass[aps,prl,twocolumn,showpacs]{revtex4}

\usepackage{graphicx}
\usepackage{bm}

\newcommand{\eps} {\varepsilon}

\begin{document}

\title{Discontinuous liquid rise in capillaries with nonuniform
cross-sections}

\author{Yoav Tsori}
\affiliation{Department of Chemical Engineering, Ben-Gurion University of the
Negev, \\P.O. Box 653, 84105 Beer-Sheva, Israel}

\date{\today}

\begin{abstract}

We consider theoretically liquid rise against gravity in capillaries with
height-dependent cross-section. For a conical capillary made from a
hydrophobic surface and dipped in a liquid reservoir, the equilibrium
liquid height depends on the cone opening angle $\alpha$, the
Young-Dupr\'{e} contact angle $\theta$, the cone radius at the
reservoir's level $R_0$ and the capillary length $\kappa^{-1}$. As
$\alpha$ is increased from zero, the meniscus' position changes
continuously until, when $\alpha$ attains a critical value, the meniscus
jumps to the bottom of the capillary. For hydrophilic surfaces the
meniscus jumps to the top. The same liquid height discontinuuity can be
achieved with electrowetting with no mechanical motion. Essentially the
same behavior is found for two tilted surfaces. We further consider
capillaries with periodic radius modulations, and find that there are few
competing minima for the meniscus location. A transition from one to
another can be performed by the use of electrowetting. The phenomenon
discussed here may find uses in microfluidic applications requiring the
transport small amounts of water ``quanta'' (volume$<1$ nL) in a regular
fashion.

\end{abstract}


\maketitle

The behavior of liquids confined by solid surfaces is important in areas
such as microfluidics \cite{pgg_b_q,pgg}, wetting of porous media
\cite{marmur1}, the creation of hydrophobic surfaces
\cite{quere1,quere2}, oil recovery \cite{mason} and water transport in
plants \cite{zimmermann}. As the system size is reduced, the interfacial 
tensions become increasingly important in comparison to bulk energies,
and are essential in understanding the equilibrium states as well as the
system dynamics.

Wetting has been studied for liquids in contact with curved surfaces
\cite{pgg_b_q,quere3,quere4,oron}, wedges \cite{finn,hauge} and cones
\cite{rejmer,parry1,parry2}, and topographically \cite{lipow1} or
chemically modulated substrates \cite{rachel1,hermin,lipow2,lipow3}.
However, surprises appear even for very simple geometries of the bounding
surfaces. Here we focus on the rise of a liquid in capillaries with
nonuniform cross-sections. When a solid capillary is immersed in a bath
of liquid, the height of the  contact line above the bath level $h$ is
given by
\begin{eqnarray}\label{jurin}
h=c\kappa^{-2}\cos\theta/R
\end{eqnarray} 
where $\kappa^{-1}\equiv (\sigma/g\rho)^{1/2}$ is the capillary length,
$\sigma$ is the liquid-gas interfacial tension, $\rho$ is the liquid mass
density (gas density neglected), $g$ is the gravitational acceleration,
and  $\theta$ is the Young-Dupr\'{e} contact angle given by
$\cos\theta=(\gamma_{gs}-\gamma_{ls})/\sigma$, where $\gamma_{gs}$ and
$\gamma_{ls}$ are the gas-solid and liquid-solid interfacial tensions,
respectively \cite{pgg_b_q}. The constant $c$ is $c=2$ for a cylindrical
capillary, in which case $R$ is the radius, or $c=1$ for two parallel and
flat surfaces separated at distance $2R$. The liquid is sucked upwards 
if the capillary's surface is hydrophilic ($\theta<\frac12\pi$), and is
depressed downwards in the case of a hydrophobic surface
($\theta>\frac12\pi$) \cite{footnote}.

\begin{figure}[h!]
\begin{center}
\includegraphics[scale=0.45,bb=90 300 510 780,clip]{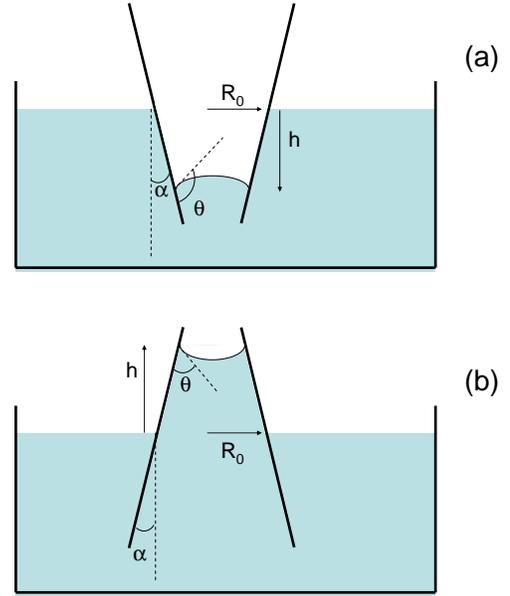}
\end{center}
\caption{Schematic illustration of cone capillary or two
tilted planes,
and definitions of parameters. Two of the possible cases: (a) hydrophobic
surface, $\cos\theta<0$, positive opening angle $\alpha$ and negative $h$. (b)
hydrophilic surface, $\cos\theta>0$, negative
$\alpha$ and positive $h$.}
\end{figure}

Suppose now that the capillary walls are not vertical, but rather have
some opening angle $\alpha$ as is illustrated in Fig. 1. What is the
liquid rise then? One can naively expect that if $\alpha$ is small, $h$
changes from Eq. (\ref{jurin}) by a small amount proportional to
$\alpha$; it is not even a-priori clear whether $h$ increases or
decreases.  We restrict ourselves to narrow capillaries, where $\kappa
R\ll 1$ is satisfied. In this case, as will be verified below, the height
is larger than the radius, $h\gg R$, and the height variations of the
meniscus surface are negligible compared to the total height.

In mechanical equilibrium, at the contact line the Laplace pressure 
is balanced by the hydrostatic pressure
\begin{eqnarray}
P_0+\frac{c\sigma}{r}=P_0-\rho gh
\end{eqnarray} 
where $P_0$ is the ambient pressure and $r$ is the 
inverse curvature, and is given by $r(h)=-R(h)/\cos(\theta+\alpha)$.
We denote $R_0$ as the radius at the bath level (see Fig. 1), and hence
$R(h)=R_0+h\tan\alpha$.
\begin{figure}[h!]
\begin{center}
\includegraphics[scale=0.75,bb=120 185 430 595,clip]{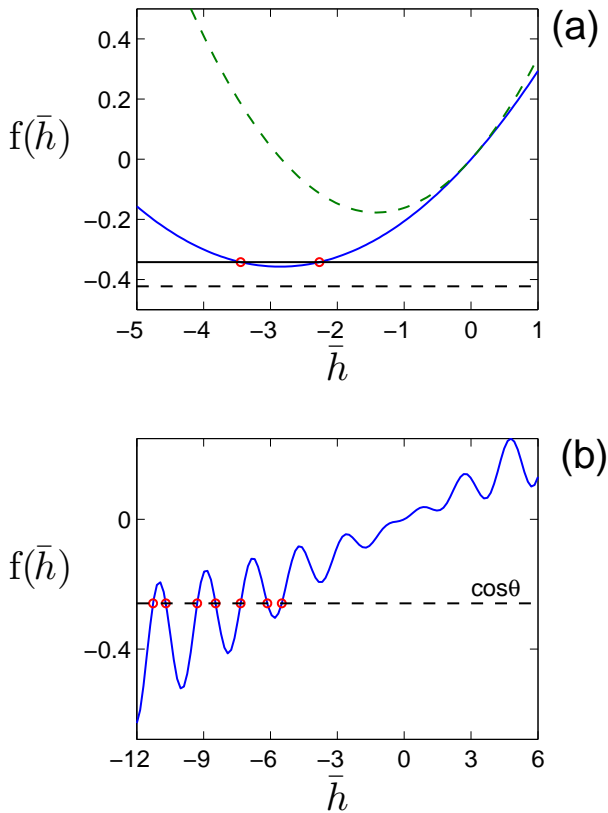}
\end{center}

\caption{(a) Solid curved line is a plot of $f(\bar{h})$ from Eq.
(\ref{gov_eqn2}) and solid horizontal line is $\cos(\theta+\alpha)$.
Their intersection occurs at two points $h_1$ (the meniscus location) and
$h_2<h_1$ marked with circles. $\bar{R}_0=0.5$, $c=2$ and $\alpha=0.087$
($5^\circ$). Dashed lines are the same, but $\alpha$ is twice as large,
$\alpha=0.174$  ($10^\circ$), above the critical angle. In this case
there is no solution to Eq. (\ref{gov_eqn1}), and the meniscus jumps down
to the bottom of the capillary. (b) Plot of $f(\bar{h})$ in the case of
periodically-modulated capillary [Eq. (\ref{gov_eqn4})]. The horizontal
dashed line is $\cos\theta$, and the multiple intersections give the
possible meniscus locations. $c=2$, $\bar{R}_0=0.07$,
$\bar{R}_m=\bar{R}_0/2$ and $\mu=3$. 
}
\end{figure}

We therefore find that the liquid rise is given by
\begin{eqnarray}\label{gov_eqn1}
\cos(\theta+\alpha)&=&f(\bar{h})\\
f(\bar{h})&=&\frac{1}{c}\bar{h}\left(\bar{R}_0+\bar{h}\tan\alpha\right)
\label{gov_eqn2}
\end{eqnarray} 
where the dimensionless variable $\bar{h}\equiv \kappa h$ and
$\bar{R}_0\equiv \kappa R_0$ have been used. These equations reduce to
the familiar form Eq. (\ref{jurin}) in the limit $\alpha\to 0$. Let us
concentrate first on the case where $\alpha$ is positive and the surface
is hydrophobic,  $\cos\theta<0$ [see Fig. 1 (a)]; the results for $\alpha<0$
follow
immediately. The left hand-side of Eq. (\ref{gov_eqn1}) is then negative
for small enough values of $\alpha$, and the quadratic form of
$f(\bar{h})$ means that the two solutions $h_1$ and $h_2$ are negative,
see Fig. 2 (a). The stable solution is $h_1$ while $h_2<h_1$ is
unstable. 

If the opening angle $\alpha$ is too large, however, the minimum of
$f(\bar{h})$, attained at $\bar{h}^*=-\bar{R}_0/(2\tan\alpha)$, is
$f(\bar{h}^*)=-\bar{R}_0^2/(4c\tan\alpha)>\cos(\theta+\alpha)$, and there
is no solution. Hence, for a given value of contact angle $\theta$, the
critical value of the opening angle $\alpha_c$ is given by the condition 
$f(\bar{h}^*)=\cos(\theta+\alpha_c)$. As $\alpha$ is increased past
$\alpha_c$, the meniscus ``jumps'' all the way to the bottom of the capillary;
in the case of a nearly closed capillary this occurs at $\bar{h}=2\bar{h}^*$.

\begin{figure}[h!]
\begin{center}
\includegraphics[scale=0.75,bb=145 180 410 620,clip]{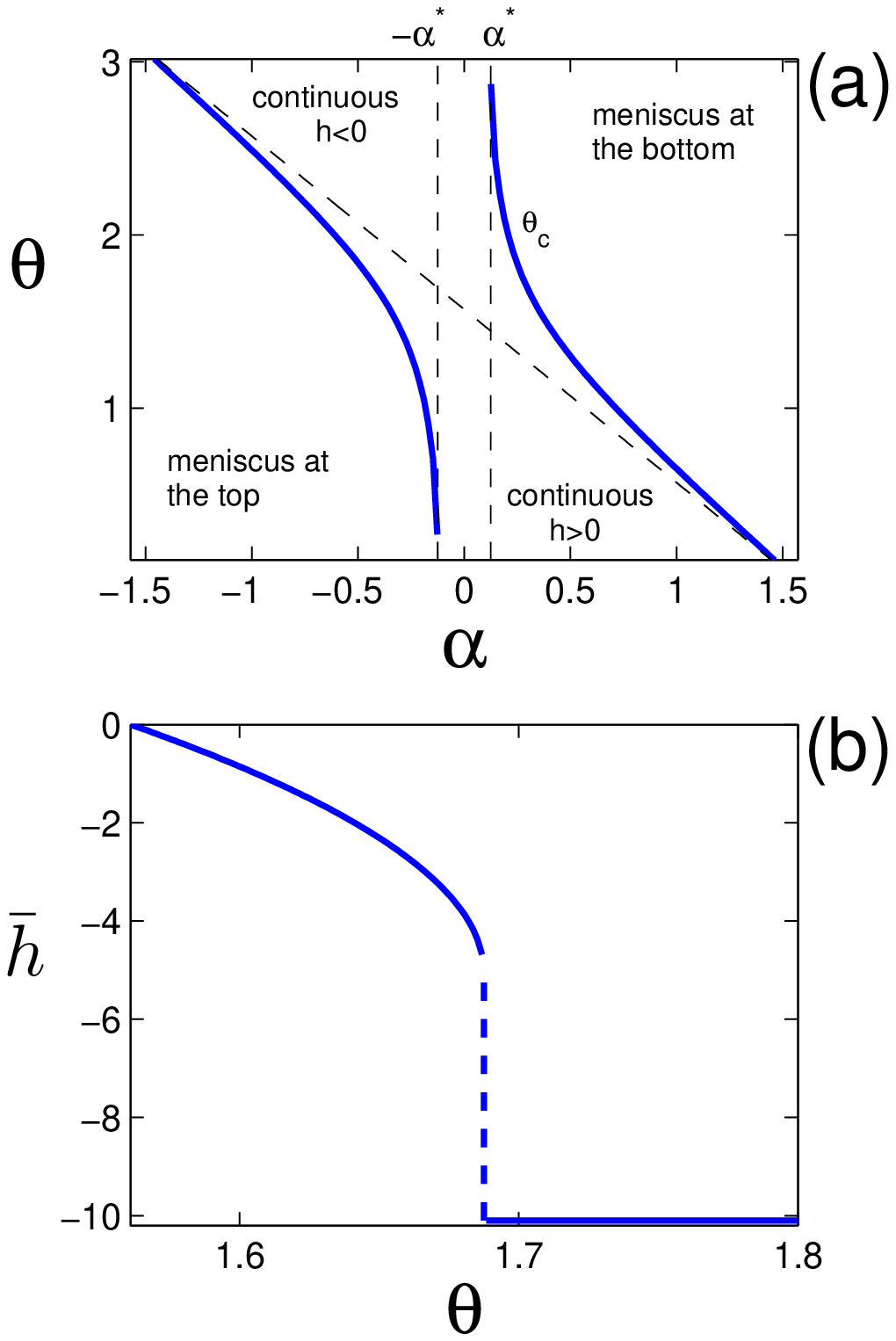}
\end{center}
\caption{(a) Phase-diagram in the opening angle-contact angle
($\alpha$-$\theta$) plane. For positive $\alpha$, solid line is
$\theta_c(\alpha)$ from Eq. (\ref{thetac}), separating two regions: below
it, the meniscus height $\bar{h}$ changes continuously as a function of
$\alpha$ and $\theta$.  $\bar{h}$ is positive if
$\frac12\pi-\alpha>\theta$, negative if
$\frac12\pi-\alpha<\theta<\theta_c$, and zero when
$\theta+\alpha=\frac12\pi$ (dashed diagonal line). At a given opening
angle $\alpha$, increase of $\theta$ past $\theta_c$ leads to a jump down
of the meniscus from $\bar{h}=\bar{h}^*<0$ to the bottom of the
capillary. Above the critical line the meniscus is
at the bottom. For all angles $\alpha<\alpha^*$ (see text) the behavior
is continuous. The phase-diagram is symmetric with respect to
$\alpha\to-\alpha$, $\theta\to\pi-\theta$ and $\bar{h}\to-\bar{h}$. 
Parameters are $\bar{R}_0=1$ and $c=2$. (b) Meniscus location $\bar{h}$
as a function of the contact angle $\theta$ at fixed $\alpha=0.01$, $c=2$
and $\bar{R}_0=0.1$. 
}
\end{figure}

When the surface is hydrophilic and both $\theta$ and $\alpha$ are small,
then there is always a positive solution for $\bar{h}$. However, if
$\theta<\frac12\pi$ but $\theta+\alpha>\frac12\pi$, the liquid height is
negative and the jump again is possible. In essence the capillary behaves
as a hydrophobic surface. 

A different approach, potentially useful in applications, is that of
electrowetting. In the experimental setup the opening angle $\alpha$ is
fixed, but the contact angle may be changed with an external potential
$V$ imposed on the walls: $\theta=\theta(V)$. The change to $\cos\theta$
is $\eps V^2/(2\sigma\lambda_D)\sim 0.3V^2$ (where
$V$ is in Volt) \cite{berge1,berge2,baret},
and thus can be quite large (we took the dielectric constant of water and
the Debye screening length $\lambda_D=10$ nm). At a fixed value of
$\alpha$, an increase in $\theta$ lowers the liquid height until $\theta$
reaches $\theta_c$ given by
\begin{eqnarray}\label{thetac}
\theta_c=\arccos\left(-\bar{R}_0^2/(4c\tan\alpha)\right)-\alpha
\end{eqnarray}
At all $\theta>\theta_c$ the meniscus jumps again to the
bottom of the capillary. However, if $\alpha<\alpha^*$, where $\alpha^*$ is
given by 
\begin{eqnarray}\label{al_st}
\sin\alpha^*=\bar{R}_0^2/4c~~~~,
\end{eqnarray} 
the liquid height is a continuous function of $\theta$ at all $\theta$.
The threshold angle $\alpha^*$ is quite small; if $\bar{R}_0=0.1$ we find
$\alpha^*=1.25\cdot 10^{-3}$ ($0.07^\circ$).

Figure 3 (a) is a phase-diagram in the $\alpha$-$\theta$ plane. In the
region marked ``continuous'' and for positive $\alpha$,
$\bar{h}(\alpha,\theta)$ changes continuously. Across the critical line
$\theta_c(\alpha)$ [Eq. (\ref{thetac})], $\bar{h}$ changes
discontinuously (meniscus is at the bottom of the capillary). Figure 3
(b) shows the height $\bar{h}$ as a function of $\theta$ at fixed value
of $\alpha$. The meniscus height $\bar{h}$ decreases below zero until, at
the critical value of $\theta$, its height jumps from $\bar{h}^*$ to the
capillary bottom (at $2\bar{h}^*$ if the capillary is nearly closed). Further
increase of $\theta$ does not change the meniscus'
location.
\begin{figure}[h!]
\begin{center}
\includegraphics[scale=0.45,bb=95 365 500 785,clip]{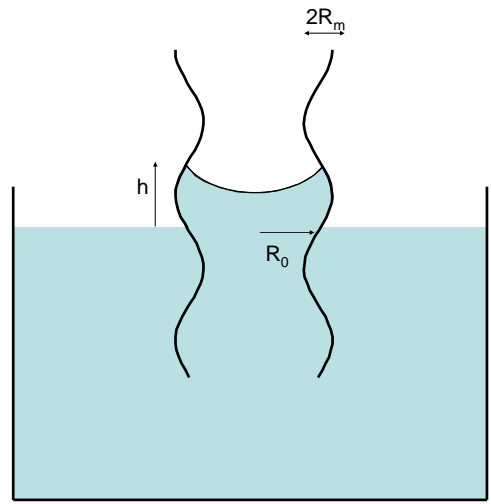}
\end{center}
\caption{Schematic illustration of a capillary with sinusoidal modulations of
the radius. 
}
\end{figure}

The liquid behavior in capillaries with negative $\alpha$ [Fig. 1 (b)]
follows from the
symmetry of the problem: the transformation $\alpha\to-\alpha$ and 
$\theta\to\pi-\theta$ leaves Eqs. (\ref{gov_eqn1}) and (\ref{gov_eqn2})
unchanged if $\bar{h}\to-\bar{h}$. 
For negative values of $\alpha$, a decrease of
$\theta$ from large values to small ones past $\theta_c$ leads to a jump of
the meniscus to the top of the capillary.

The above insight can be used to exploring different capillaries, and we
briefly mention a capillary with periodic width modulations
\cite{rungta,ross}, namely $R(h)=R_0+R_m\sin(qh)$, where $R_m$ is the
modulation amplitude and $q$ its wavenumber. We restrict ourselves to the
long wavelength regime, where $qR_m\ll 1$. In this case it can be shown
that the governing equations replacing Eqs. (\ref{gov_eqn1})  and
(\ref{gov_eqn2}) are
\begin{eqnarray}\label{gov_eqn3}
\cos\theta &=& f(\bar{h})\\
f(\bar{h}) &=& \frac{1}{c}\bar{h}\left(\bar{R}_0+
\bar{R}_m\sin(\mu\bar{h})\right)\label{gov_eqn4}
\end{eqnarray} 
where $\bar{R}_m\equiv\kappa R_m$ and $\mu=q/\kappa$. It is clear 
from Fig. 2 (b) that there are multiple solutions, half of which are
maxima and the other half minima. 

For a system prepared in a given minimum, increasing $\theta$ by the use
of electrowetting decreases $\cos\theta$. Thus, the liquid location
changes -- $h$ decrease. When the liquid height overlaps with a minimum
of $f(\bar{h})$, further increase of $\theta$ leads to a jump in the
liquid height to the next ``branch'' of $f(\bar{h})$. In this way one
``quantum'' of liquid is depleted from the capillary; if $\theta$ is
decreased, at each step one liquid unit is sucked into the capillary. The
unit volume can be estimated to be $v\sim R_0^2/q$; for a capillary width
of $R_0=100$ $\mu$m and wavenumber $q=10^3$ m$^{-1}$, we find $v=10$ nL,
whereas reducing the sizes to $R_0=10$ $\mu$m and $q=10^4$ m$^{-1}$ gives
$v=10^{-2}$ nL.

In summary, we have shown that the liquid height in capillaries with
nonuniform cross-sections is a discontinuous function of the geometrical
variables. This peculiar phase-transition is important for the
understanding of liquids confined to small environments, as capillaries
in practice rarely  have uniform cross-sections. Indeed, since $\alpha^*$
can be extremely small, if the surfaces of the capillary are
super-hydrophobic, very small deviations of $\alpha$ around $\alpha=0$
will yield discontinuous liquid heights. It may be beneficial to exploit
the dependence of the water level on the contact angle (e.g.) in
microfluidic applications where it is desired to accurately control small
volumes of liquid. The setup of Fig. 1 could possibly be used as a
``switch'' to prevent or allow liquid flow or electrical current in the
direction perpendicular to the plane of the paper, while the setup of
Fig. 4 permits to ``suck'' known quantities of fluids.

In this paper we assumed ideal surfaces and thus have not dealt with
rough surfaces and hysteresis effects \cite{reed}. It would also be
interesting to consider a pressurized capillary (micro-pipette), where
the finite pressure difference between the capillary and the ambient
atmosphere can bring about further surprises \cite{in_prep}.


\begin{acknowledgments}
{\bf Acknowledgments} 
I am indebted to P. -G. de Gennes with whom I had numerous fruitful
discussions
and correspondences on the subject, and who has dearly helped me during my
stay
in France.
For stimulating discussions and corrections to
the manuscript I would like to thank F. Brochard-Wyart, H. Diamant, A.
Marmur, I.
Szleifer, and R. Yerushalmy-Rozen. 
This research was supported in part by the Israel Science foundation (ISF)
grant number 284/05.
\end{acknowledgments}



\begin{thebibliography}{0}



\bibitem{pgg_b_q} {\it Gouttes, bulles, perles et ondes}, P. -G. de Gennes,
F. Brochard-Wyart and D. Qu\'{e}r\'{e} (Belin 2002).

\bibitem{pgg} P. -G. de Gennes, {\it Rev. Mod. Phys.} {\bf 57}, 827 (1985).

\bibitem{marmur1} 
A. Marmur, {\it J. Colloid. Interf. Sci.} {\bf 129}, 278 (1989); A.
Marmur and R. D. Cohen, {\it J. Colloid. Interf. Sci.} {\bf 189}, 299 (1997).

\bibitem{quere1} J. Bico, C. Marzolin and D. Qu\'{e}r\'{e}, {\it Eur. Phys.
Lett.} {\bf 47}, 220 (1999).

\bibitem{quere2} A. Lafuma and D. Qu\'{e}r\'{e}, {\it Nature Mater.} {\bf 2},
457 (2003).

\bibitem{mason} N. R. Morrow and G. Mason, {\it Curr. Opin. Colloid. Interface.
Sci.} {\bf 6}, 321 (2001).

\bibitem{zimmermann} M. H. Zimmermann, {\it Scient. Am.} {\bf 208}, 133 (1963).

\bibitem{quere3} D. Qu\'{e}r\'{e}, J. M. Di Meglio and F. Brochard-Wyart, {\it
Science} {\bf 249}, 1256 (1990).

\bibitem{quere4} D. Qu\'{e}r\'{e} and J. M. Di Meglio, 
{\it Colloid. Interf. Sci.} {\bf 48}, 141 (1994).

\bibitem{oron} A. Oron, S. H. Davis and S. G. Bankoff, {\it Rev. Mod. Phys.}
{\bf 69}, 931 (1997).

\bibitem{finn} 
P. Concus and R. Finn, {\it Proc. Natl. Acad. Sci. USA} {\bf 63}, 292
(1969).

\bibitem{hauge} 
E. H. Hauge, {\it Phys. Rev. A} {\bf 46}, 4994 (1992).

\bibitem{rejmer} K. Rejmer, S. Dietrich and M. Napi\'{o}rkowski,
{\it Phys. Rev. E.} {\bf 60}, 4027 (1999).

\bibitem{parry1} 
A. O. Parry, C.
Rascon and A. J. Wood, {\it Phys. Rev. Lett.}  {\bf 83}, 5535 (1999).

\bibitem{parry2} 
A. O. Parry, A. J. Wood and C. Rascon, {\it J.
Phys.: Cond. Mat.} {\bf 13}, 4591 (2001).


\bibitem{lipow1} R. Seemann, M. Brinkmann, E. J. Kramer, F. F. Lange and R.
Lipowsky, {\it Proc. Natl. Acad. Sci. USA} {\bf 102}, 1848 (2005).

\bibitem{rachel1} 
J. Jopp, H. Gr\"{u}ll and R.
Yerushalmi-Rozen, {\it Langmuir} {\bf 20}, 10015 (2004).

\bibitem{hermin} G. Hartmut, S. Herminghaus, P. Lenz and R. Lipowsky,
{\it Science} {\bf 283}, 46 (1999).

\bibitem{lipow2} M. Brinkmann and R. Lipowsky, {\it J. Appl. Phys.} {\bf 92},
4296 (2002).

\bibitem{lipow3} R. Lipowsky, {\it Curr. Opin. Colloid. Interf. Sci.} {\bf 6},
40 (2001).

\bibitem{footnote} In the above and throughout this paper, line tension effects
are ignored because the length-scales considered are not small enough
\cite{pgg_b_q}; these effects can be incorporated as well
\cite{lin_lib,jensen_li}.

\bibitem{lin_lib} 
F. Y. H. Lin and D.
Lib, {\it Colloids and Surfaces A: Physicochemical and Engineering
Aspects} {\bf 87}, 93 (1994).

\bibitem{jensen_li} 
W. C. Jensen and D. Q. Li, {\it
Colloids and Surfaces A-Physicochemical and Engineering Aspects} {\bf
156}, 519 (1999).

\bibitem{berge1} C. Quilliet and B. Berge, {\it Eur. Phys. Lett.} {\bf 60},
 99 (2002).

\bibitem{berge2} C. Quilliet and B. Berge, {\it Curr. Opin. Colloid. Interf.
Sci.} {\bf 6}, 34 (2001).

\bibitem{baret} F. Mugele and J.-C. Baret, {\it J. Phys.: Condens. Matter.}
{\bf 17}, R705 (2005).

\bibitem{rungta} 
A. Borhan and K. K. Rungta, {\it J. Colloid. Interf. Sci.} {\bf 155}, 
438 (1993).

\bibitem{ross} 
R. Sharma and D. S. Ross, {\it J. Chem. Soc. Faraday Trans.}
{\bf 87}, 619 (1991).

\bibitem{reed} S. Levine, J. Lowndes and P. Reed, {\it J. Colloid. Interf.
Sci.} {\bf 77}, 253 (1980).

\bibitem{in_prep} Y. Tsori, manuscript in preparation (2006).



\end{thebibliography}
\end{document}